\documentclass[aps,amsmath,amssymb,prl,twocolumn,showpacs,superscriptaddress]{revtex4}
\usepackage{bm}
\usepackage{epsfig}

\begin{document}

\title{$1/f$ Scaling in Heart Rate Requires Antagonistic Autonomic Control}

\author{Zbigniew R. Struzik}
\email{z.r.struzik@p.u-tokyo.ac.jp}
\affiliation{Educational Physiology Laboratory, Graduate School of
  Education, The University of Tokyo, 7--3--1 Hongo, Bunkyo-ku, Tokyo
  113--0033, Japan}
\affiliation{PRESTO, Japan Science and Technology Agency, Kawaguchi,
  Saitama 332--0012, Japan}
\author{Junichiro Hayano}
\author{Seiichiro Sakata}
\affiliation{Core Laboratory, Nagoya City University Graduate School
  of Medical Sciences, 1 Kawasumi, Mizuho-cho, Mizuho-ku, Nagoya
  467--8601, Japan}
\author{Shin Kwak}
\affiliation{Department of Neurology, Graduate School of Medicine,
  The University of Tokyo, 7--3--1 Hongo, Bunkyo-ku, Tokyo 113--0033,
  Japan}
\author{Yoshiharu Yamamoto}
\email{yamamoto@p.u-tokyo.ac.jp}
\affiliation{Educational Physiology Laboratory, Graduate School of
  Education, The University of Tokyo, 7--3--1 Hongo, Bunkyo-ku, Tokyo
  113--0033, Japan}
\affiliation{PRESTO, Japan Science and Technology Agency, Kawaguchi,
  Saitama 332--0012, Japan}

\date{\today}

\begin{abstract}
We present the first systematic evidence for the origins of $1/f$-type
temporal scaling in human heart rate. The heart rate is regulated by the
activity of two branches of the autonomic nervous system: the
parasympathetic (PNS) and the sympathetic (SNS) nervous systems. We
examine alterations in the scaling property when the balance
between PNS and SNS activity is modified, and find that the
relative PNS suppression by congestive heart failure results in a
substantial increase in the Hurst exponent $H$ towards random walk
scaling $1/f^{2}$ and a similar breakdown is observed with relative
SNS suppression by primary autonomic failure. These results suggest
that $1/f$ scaling in heart rate requires the intricate balance between
the {\it antagonistic\/} activity of PNS and SNS.
\end{abstract}

\maketitle

Healthy human heart rate has long been known to exhibit $1/f$-type
fluctuations \cite{656,1357,766,1430} and has recently also been
attributed multifractal scaling properties \cite{1620}. This
complex dynamics, resembling non-equilibrium \cite{1522} and/or
multi-scale \cite{1759} dynamics in physics, has been demonstrated
to be independent of human behavior---the statistical properties
of heart rate remain unaltered even after eliminating known
behavioral modifiers \cite{Amaral01,Aoyagi03}---, suggesting that
the origin of heart rate complexity lies in the intrinsic dynamics
of the physiological regulatory system. One conjecture previously
posed is that $1/f$ (global) scaling and local multifractal scaling
in heart rate is caused by the interaction between the activity of
sympathetic (SNS) and parasympathetic (PNS) nervous systems \cite{1357},
leading respectively to the increase and the decrease in heart rate.
However, the evidence for this is scarce \cite{attempt}.

Here, we present the first systematic evidence for the origins of
$1/f$ scaling and multifractality in human heart rate. We
demonstrate that modifying the relative importance of either of
the two branches leads to a substantial departure from $1/f$ scaling,
showing that $1/f$ scaling in healthy heart rate requires the
existence of and the intricate balance between the antagonistic activity
of PNS and SNS. It supports the view of the cardiac neuroregulation
as a system in a critical state \cite{Bak87}, and permanently out of
equilibrium, in which concerted interplay of the SNS and PNS is
required for preserving momentary ``balance''. This view of cardiac
neuroregulation is coherent with a broad class of models of phenomena
which, to a large extent, has been established using the implicit
or explicit concept of balance of competing agents or scenarios.

Further, we also observe a hitherto unexplored relationship
between the multifractality of the heart rate and variability as
measured by interval variance.  While it is generally believed that
lower variability results in a reduction of multifractal properties
(reduced spectrum width), as has been demonstrated in relative PNS
suppression both by congestive heart failure (CHF) \cite{1620} and by
the parasympathetic blocker {\it atropine\/} \cite{Amaral01}, we
observe conservation of multifractal properties in relative SNS
suppression by primary autonomic failure (PAF) at substantially
reduced variability to levels closer to CHF. This suggests the
relevance of the intrinsic PNS dynamics for multifractality. 

We believe these findings to be important in putting forward the
{\it antagonistic scenario\/} for complex (multi-) fractal dynamics
that has now been observed in a wide variety of real-world signals
and also in helping diagnose the condition of a range of patients
having abnormality in their autonomic regulatory system.

\begin{figure}[t]
\vspace{0.03in}
\hbox{
\hspace{-0.12in}
\includegraphics[width=1.03\linewidth]{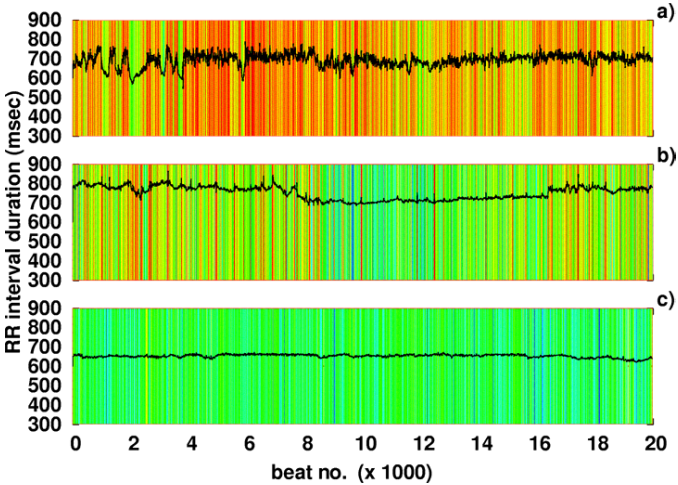}
}
\caption{(Color) Typical traces of daytime heartbeat intervals for (a) 
  a healthy subject, (b) a PAF patient and (c) a CHF patient. 
  The color coding used shows the local contribution to
  multifractality---the color spectrum is centered at $h=0.3$ (green),
  with the strongest singularities in red ($h=0.0$) and the weakest
  in blue ($h=0.6$) \cite{Fractals00}.}
\label{fig:fig1}
\end{figure}

We analyze three groups of subjects, of whom long-term heart rate
data were measured as sequential heart interbeat intervals.
The first group consists of 115 healthy subjects ($26$ women and
$89$ men; ages $16-84$ yrs) without any known disease affecting
autonomic controls of heart rate, who underwent ambulatory
monitoring during normal daily life [Fig.~\ref{fig:fig1}(a)].
The total number of whole-day data sets is $181$, as some of the
subjects were examined for two consecutive days, with each data
set containing on average $10^{5}$ heartbeats. Details of the
recruitment of the subjects, screening for medical problems, protocols
and the data collection are described in Ref.~\cite{Sakata99}.
We analyzed both whole-day data containing sleep and awake
periods and daytime only data, with essentially identical
results \cite{note_day_night}. In this paper, we present daytime
results only. 

The second group of subjects are 12 patients with CHF, of whom
whole-day ambulatory data [Fig.~\ref{fig:fig1}(c)] are available from
Physionet \cite{Physionet}. This severe heart failure is known to
be associated with both increased SNS \cite{1314,Elam03} and decreased
PNS \cite{1314,773} activity. Thus, this data set contains
information on how heart rate is (multi-) scaled during relative
PNS suppression.

As the last group, we examined the 24-hour ambulatory heart rate
dynamics of 10 PAF patients aged 54--77 years \cite{PAF,Oppenheimer80},
containing on average $10^{5}$ heartbeats \cite{hrv_measure}
[Fig.~\ref{fig:fig1}(b)]. PAF is clinically characterized as
autonomic dysfunction, including orthostatic hypotension, impotence,
bladder and bowel dysfunction and sweating defects, which 
primarily result from progressive neuronal degeneration of unknown
cause. The main pathological finding related to autonomic dysfunction
in PAF is severe loss of preganglionic and/or postganglionic
sympathetic neurons \cite{Matthew99}. In contrast to the severe 
degeneration of the efferent SNS, PNS is believed to remain
relatively intact in PAF; we will confirm this below by showing
a similar level of high-frequency fluctuations of heart rate,
known as a robust indicator of PNS activity \cite{1099,865}, in our
PAF patients to that of healthy subjects. Thus, it is highly possible 
that this group serves as an example of relative and {\it neurogenic\/}
SNS suppression. 

The mean global scaling exponent (the Hurst exponent $H$) has been
evaluated by using (first order) detrended fluctuation analysis (DFA)
\cite{Peng93,Peng95,Physionet}. 
We have analyzed the scaling behavior of the mean quantity (group mean)
$\bar{M}_{DFA}(s)=L^{-1}\sum_{l=1}^{L}\log_{10}(D^{(l)}_{DFA}(s))$,
where $l$ 
indexes time series in the group.
For each scale/resolution $s$ as measured by the DFA window size, and
for each integrated, normalized heartbeat interval time series
$\{F^{(l)}_i={T_l^{-1}}\sum_{j=1}^i{f^{(l)}_j}\}_{(i=1, \dots, N_l),(l=1, \dots, L)}$,
$D^{(l)}_{DFA}(s)$ (total scalewise detrended fluctuation) has been
calculated: 
\[D^{(l)}_{DFA}(s)={s^{-1}}\sqrt{ \frac{1}{K^{(l)}(s)} \sum_{k=1}^{K^{(l)}(s)}(F^{(l)}_k-P_k(s))^2}\;.\]
$P_k(s)$ denotes the local least-squares linear fit in each DFA window $k$, 
and $K^{(l)}(s)$ is the number of windows per scale $s$. 
Integration of the input heart rate intervals is performed according
to standard DFA practice, and the norm used is the elapsed time
$T_l=\sum_{i=1}^{N_l}f_i^{(l)}$. 
The normalization applied allows us to compute group averages of 
records of different duration, and to compare the mean absolute 
levels of variability per resolution $s$; for each resolution $s$, the
quantity $\bar{M}_{DFA}(s)$ measures the (logarithmic) scalewise mean 
of the normalized DFA---the sum of the logarithm of detrended 
fluctuations for each group of time series at this resolution. 
The Hurst exponent was computed from the log-log 
fit to the group averaged DFA
values over the selected range of scales ($20-4,000$ beats).

\begin{figure}[t]
\vspace{-0.04in}
\hbox{
\hspace{-0.31in}
\includegraphics[width=1.107\linewidth]{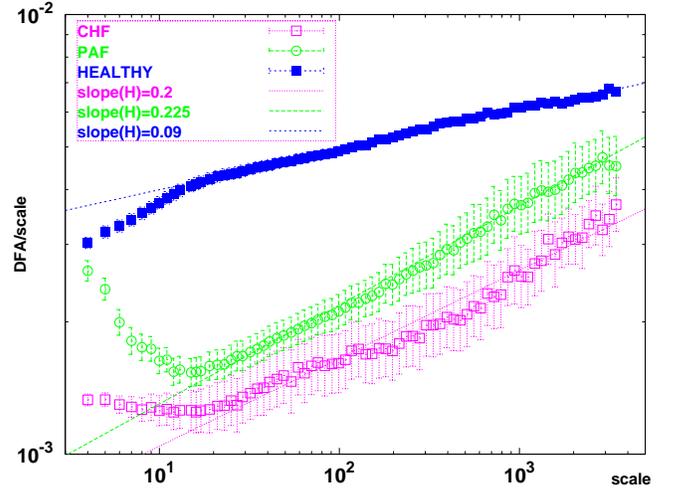}
}
\vspace{-0.03in}
\caption{(Color online) Scale dependency of the mean detrended fluctuation
  $\bar{M}_{DFA}(s)$ for healthy subjects, PAF patients and CHF patients.
  Detrended fluctuations have been calculated with first order DFA, i.e.
  linear trend removal \cite{Peng95}. Vertical bars represent the
  standard deviations of the group means.} 
\label{fig:fig2}
\end{figure}

In Fig.~\ref{fig:fig2}, we show the scaling behavior of the
$\bar{M}_{DFA}(s)$ 
versus $\log_{10}(s)$ for healthy
subjects, PAF patients, and CHF patients, with the slopes
corresponding to the Hurst exponent $H$.
We find a substantial difference in the scalewise variability
levels between controls and PAF and CHF patients. This holds for
the entire compared resolution range of $4-4,000$ beats as measured
by the DFA window size $s$. However, PAF variability reaches normal
levels asymptotically for the highest resolutions (and lowest beat numbers),
most likely reflecting the preservation of high-frequency
fluctuations of heart rate indicative of the intact PNS activity
\cite{1099,865} in our PAF patients. CHF variability, on the contrary,
remains at low levels at all resolutions. 

In addition, we find that the relative PNS suppression by CHF 
results in a substantial increase in the Hurst exponent from $1/f$
range ($H\approx 0.09$ for healthy controls) to $H>0.2$, i.e. towards
random walk scaling $1/f^{2}$ ($H=0.5$) [Fig.~\ref{fig:fig2}]. This
effect has been observed for the entire range of resolutions with
almost consistent scaling, which for all three groups stretches from
about 20 beats up to the maximum resolution used of 4,000 beats
(DFA window size). The slope within the scaling range obtained for
PAF is close to that obtained for CHF and considerably higher than
that for the control group. Thus, surprisingly, we observe a similar
breakdown in the case of relative and neurogenic SNS suppression by
PAF. This is particularly interesting in the context of the recognized
effect that $\beta$-adrenergic blockers, mainly affecting the response
of the heart to non-suppressed SNS activity and leaving the vascular
branch of sympathetic neuroregulation intact, do not result in a
breakdown of $1/f$ scaling in heart rate \cite{Amaral01,1430}.

\begin{figure}[t]
\hbox{
\hspace{-0.30in}
\includegraphics[width=1.107\linewidth]{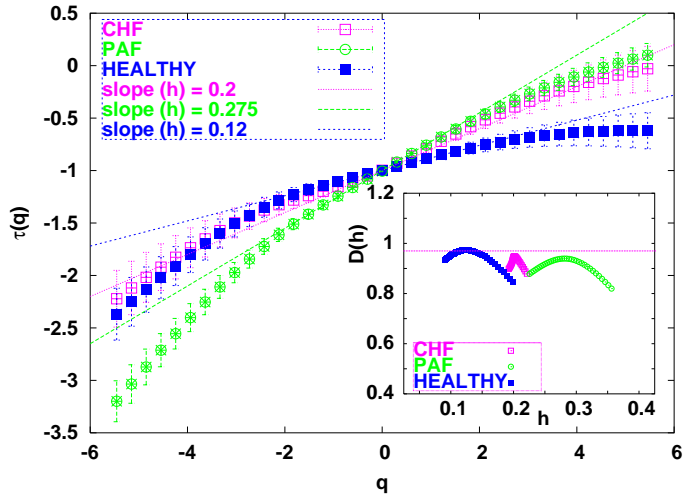}
}
\caption{(Color online)
 Multifractal ($\tau(q)$) spectra for healthy subjects, PAF patients,
 and CHF patients. 
Vertical bars represent the standard deviations of the linear fit to
 (the group mean of) the partition function. (Inset) Singularity ($D(h)$)
 spectra derived from the avarage $\tau(q)$ curves.}
\label{fig:fig3}
\end{figure}

Further, we also tested the multifractal properties of the
data using the wavelet-based multifractal methodology \cite{1758}.
We apply the 2nd derivative of the Gaussian to the data as the
mother wavelet before calculating the partition function $Z_q(s)$,
defined as the sum of the $q$-th powers of the local maxima of the
modulus of the wavelet transform coefficients at scale $s$.
The power law scaling of $Z_q(s)$ for $13 < s < 850$ then yields the
scaling exponents $\tau(q)$---the multifractal spectrum
[Fig.~\ref{fig:fig3}]. The multifractal spectrum is related to the 
singularity spectrum $D(h)$, where $D(h_o)$ is the fractal dimension
of the subset of the original time series characterized by a
{\it local\/} Hurst exponent $h=h_o$ \cite{Vicsek}, through a Legendre
transform $D(h) = q h - \tau(q)$ with $h = d\tau(q) / d q$
[Fig.~\ref{fig:fig3}, inset].

For both PAF patients and control subjects, we obtained comparable
curvature of the $\tau(q)$. However, this curvature is nearly lost
in the case of CHF patients [Fig.~\ref{fig:fig3}]. These results
imply wider singularity spectra $D(h)$ for both PAF patients and control
subjects, indicative of preserved multifractality [Fig.~\ref{fig:fig3},
inset], that can also be observed in non-uniform distributions of
the local Hurst exponents $h$ [Fig.~\ref{fig:fig1}]. In addition, we
also observe an intriguing relation between the conserved
multifractality of the heart rate for the PAF case and the
profoundly low absolute variability as measured by normalized DFA
[Fig.~\ref{fig:fig2}]. While it is generally believed that lower
variability results in a reduction of multifractal properties (reduced
spectrum width), as has been demonstrated in relative PNS suppression
both by CHF \cite{1620} and the parasympathetic blocker {\it atropine\/}
\cite{Amaral01}, in PAF patients we observe conservation
of multifractal properties at substantially reduced variability to the
levels closer to CHF. This suggests the relevance of the intrinsic PNS
dynamics for the multifractality of heart rate.

Amaral {\it et al.\/} \cite{Amaral01} reported a slightly decreased
width of multifractal spectra and an almost unchanged global scaling
exponent of healthy heart rate during the administration of the
sympathetic blocker {\it metoprolol\/}, which reduces sympathetic
control by blocking the action of $\beta$-adrenergic receptors on the heart.
Physiologically, this case is very different from PAF because in
healthy subjects, central SNS activity also influences vascular tone
of both capacitance and resistance vessels through $\alpha$-adrenergic
mechanisms, and strongly affects blood pressure (through cardiac
output and peripheral resistance) and hence heart rate through
{\it baroreflex\/} control with the intact PNS. In other words,
while the {\it metoprolol\/} only blocks the $\beta$-adrenergically mediated
effects of SNS on the heart in healthy subjects, PAF is associated
with a wide range of sympathetic failure affecting various
end-organs including the heart and the vasculature \cite{Matthew99},
and the breakdown of $1/f$ scaling of heart rate is observed only
in the latter case. Thus, we conclude that healthy $1/f$ heart
rate indeed requires physiologically antagonistic activity of PNS
and SNS within the brain for autonomic neuroregulation.

A relevant question would be why ``nature'' has implemented an
antagonistic control system in one of, if not the, most important
instruments in maintaining human life, i.e. the heart. One
possible, albeit as of now still speculative explanation is that the
antagonistic control prevents mode locking by ensuring permanent
far-from-equilibrium-like, critical state-like operation \cite{1522},
and thus enhances error tolerance of the system \cite{1165}.
The importance of this invariant ``response''---mode-free operation---may
be the result of the optimization of the heart rate control system
by evolutionary processes; physiologically antagonistic cardiac
control is observed in a wide range of vertebrates \cite{Taylor99}.
A mode-free response may be important for rapid change of the
operating point of the system according to dynamically changing
internal and/or external environmental conditions.

Historically, measurements of fluctuations of heart rate have been
widely used for monitoring human autonomic controls in health and
disease \cite{1099,865}. In particular, the heart rate can easily
be measured during normal daily life, by ambulatory monitoring
devices, enabling us to probe various autonomic pathologies in a
natural setting. However, one drawback of this method using
short-term fluctuations of heart rate such as spectral analyses
\cite{1099,865} is that the statistical properties of heart rate may
be affected by behavior (e.g. exercise, diet, postural changes, etc.),
as well as by pathological changes in the autonomic nervous
system. It is usually very difficult to monitor patients' behavior
during normal daily life, and robust identification of autonomic
abnormality due to the disease per se is difficult.
By contrast, the long-term (multi-) scaling properties of ambulatory
heart rate have recently been shown to be highly independent of
behavioral modifiers \cite{Amaral01,Aoyagi03}. This study further
shows that the scaling properties do depend on the autonomic
pathologies of patients, i.e. one may be able to derive a
behavioral-independent marker for PNS suppression by the increased
global scaling exponent and the decreased multifractality of heart
rate, and for SNS suppression by the increased global exponent,
but with preserved multifractality. Thus, our findings could be also
important in helping diagnose a range of patients having abnormality
in their autonomic regulations.

We thank Dr. K. Kiyono, Prof. K. Nakahara, and Dr. S. Murayama
for their help and discussion. This study was in part supported
by Japan Science and Technology Agency.


\begin{thebibliography}{30}
\bibitem[{attempt()}]{attempt}
\bibinfo{note}{
A recent attempt to provide such evidence~\cite{Amaral01} through
drug-induced suppression study has not been fully succesful and is rather
difficult to interpret because the suppression of only one branch of
the autonomic regulatory system at the effector level (i.e. the heart)
would lead to compensatory dynamics through the other intact branch.}
\bibitem[{not()}]{note_day_night}
\bibinfo{note}{This confirms the view that sleep and other behavioral modifiers
  do not alter the scaling property of heart rate within about an hour
  \cite{Aoyagi03}, i.e. in the same range used in the current study.}
\bibitem[{PAF()}]{PAF}
\bibinfo{note}{Eight of the patients were with multiple system atrophy
  \cite{Oppenheimer80} with varying degrees of Parkinsonism, ataxia and
  pyramidal signs. The other two had pure autonomic failure
  \cite{Oppenheimer80} with no other neurological symptoms; other possible
  causes of their autonomic dysfunction were excluded by extensive laboratory
  studies. All of them performed daily activities independently. The patients
  did not take vasoactive medicines on the day they were tested. The study was
  reviewed and approved by the ethics committee of The University of Tokyo
  Hospital.}
\end{thebibliography}
\end{document}